\documentstyle[aps,psfig,multicol]{revtex}

\begin{document}

\draft

\title{A model of macroevolution with a natural system
size\footnote{Submitted to IMA Journal Mathematics Applied in Medicine
and Biology}}

\author{D A Head\footnote{Electronic address: David.Head@brunel.ac.uk}}
\address{Institute of Physical and Environmental Sciences, Brunel University,
Uxbridge, Middlesex, UB8~3PH, United Kingdom.}

\author{G J Rodgers\footnote{Electronic address: G.J.Rodgers@brunel.ac.uk}}
\address{Department of Mathematics and Statistics, Brunel University,
Uxbridge, Middlesex, UB8~3PH, United Kingdom.}

\date{14$^{\rm th}$ July, 1998.}

\maketitle

\begin{abstract}
We describe a simple model of evolution which incorporates
the branching and extinction of species lines,
and also includes abiotic influences.
A first principles approach is taken in which the probability
for speciation and extinction are defined purely in terms
of the fitness landscapes of each species.
Numerical simulations show that the total diversity fluctuates
around a natural system size $N_{\rm nat}$ which only weakly
depends upon the number of connections per species.
This is in agreement with known data for real multispecies
communities.
The numerical results are confirmed by approximate mean field
analysis.
\end{abstract}

\pacs{Keywords: Macroevolution, Bak-Sneppen model, speciation, extinction}

\maketitle

\begin{multicols}{2}
\narrowtext


The Bak-Sneppen model was introduced to illustrate the
possible role of self-organised criticality in
evolving ecosystems (Bak 1993, 1997).
It is a toy model that describes every species by a single scalar
quantity, relating to the expected time before that species evolves
to a new form.
Interactions in the ecosystem are incorporated by assuming that
a species that evolves can alter the time taken for other species
to evolve, such as those involved in predator-prey or host-parasite
relationships.
The model is said to be {\em self-organised critical}\/ because
it approaches a critical state without any apparent need for
fine parameter tuning.
Consequently, it predicts that extinction events of magnitude $\Delta E$
should occur
with a frequency proportional to $(\Delta E)^{-\alpha}$, which is
consistent with known paleobiological data (Sol\'{e} 1996).
Further evidence has come from analysis of the temporal distribution
of extinctions, which exhibits ``1/f noise'', also predicted
by the model (Sol\'{e} 1997a).

Other simple models of macroevolution have now been devised which also
claim agreement with the paleobiological
data (Peliti 1997).
Some of these are believed to be self-organised critical (Sol\'{e} 1997b),
although others exhibit power law behaviour via different mechanisms
(Roberts 1996, Newman 1997).
All of these models have in common the assumption of a constant
system size.
This has been justified by assuming that each species
occupies a single {\em ecological niche}, and that
if a species is made extinct its niche is
immediately filled by a similar, newly emerged species.
The concept of an ecological niche refers to a
set of conditions and interactions within the ecosystem
that only a single species can satisfy.
However, since the definition of a niche depends upon the other species,
the total number of niches should be defined
from within the system itself rather than being
fixed to some arbitrary constant
value for the benefit of computer simulation.

Models have been devised which are similar to the Bak-Sneppen model
but allow the total number of species to vary in time.
Kramer {\em et al.}\/ introduced a model in which the species
are placed onto a branching phylogenetic tree structure, where
each species only interacts with its closest relatives
(Kramer 1996).
Depending upon the choice of a parameter, the tree either expands
indefinitely or stops growing after a finite time.
In the model of Wilke {\em et al.} the system refills after an extinction
event at a rate according to a parameter $N_{\rm max}\,$, which is
``the maximal number of species that can be sustained with the available
resources'' (Wilke 1997).
However, since the resources are themselves biotic, we believe that
any such $N_{\rm max}$ should be defined from within the system.

In this paper, we derive and study a version of the Bak-Sneppen model
in which the probabilities for speciation and extinction are defined
purely in terms of the individual species.
Nonetheless, the total diversity fluctuates around a natural
system size without the need for global control.
In particular, there is no recourse to ``ecological niches''.
The rules of the model are based on careful considerations of the
motion of species on their fitness landscapes.
This is described in Sec.~\ref{s:se_desc}, and the results of
numerical simulations of the model are presented in Sec.~\ref{s:se_num}.
Comparisons to real ecosystems are made in Sec.~\ref{s:se_real}.
Approximate analysis of the model is presented in
Sec.~\ref{s:se_anal} which supports and enhances the numerical
work.
Finally, we discuss our results in Sec.~\ref{s:se_disc}.

\addtocounter{section}{1}
\section{Quantitative desc\-rip\-tions of macro\-evolution}
\label{s:se_desc}

\begin{figure}
\hspace{0.05in}
\psfig{file=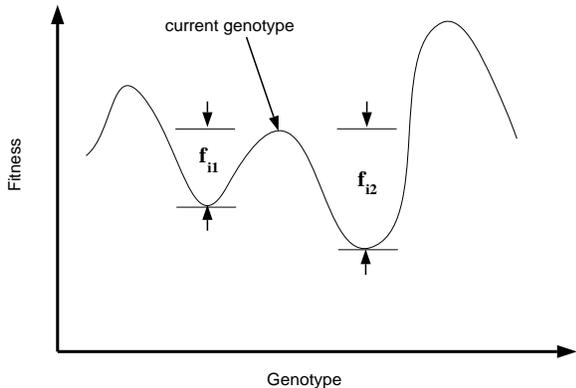,width=3in}
\caption{Schematic of a typical fitness landscape for a species labelled~$i$,
where for clarity the space of all possible genotypes
has been compressed onto a single axis. The fitness scale is arbitrary.
The barriers have heights of $f_{i1}$ and~$f_{i2}$.
}
\label{f:fit_land}
\end{figure}

To quantitatively describe an evolving ecosystem requires
some general principle that applies equally to all species.
A candidate for such a principle is to consider the
relationship between an organism's genotype and its {\em fitness},
which is some measure of the expected number of
genes passed back into the species' gene pool (Dawkins 1983).
Each point in the multidimensional space of all possible genotypes
can be assigned its own fitness value, forming a
{\em fitness landscape}, as schematised in Fig.~\ref{f:fit_land}.
The process of evolution can then be described as a
walk over this landscape in the direction of increasing fitness.
Rather than try to calculate the fitness for each genotype from
first principles, clearly a hopeless task,
Kauffman has assumed that the relationship
is so complex that it can be well approximated by random
variables (Kauffman 1993).
This leads to the concept of {\em rugged fitness landscapes},
where ``rugged'' refers to the large variations in fitness that can
result from small changes in genotype.

Models based on the Bak-Sneppen approach assume that
each species moves on its own rugged fitness landscape,
eventually becoming trapped in the region of a local maximum.
If the landscape is fixed, then the species can only evolve
by moving to a different maximum.
Suppose that a species labelled $i$ is at a local
maximum which is separated from nearby maxima
by fitness barriers of heights~$f_{ij}\,$,
where \mbox{$j=1,2,3,\ldots$} and the
$f_{ij}$ are ordered such that
\mbox{$f_{ij}\leq f_{ij+1}$}.
Over time, fluctuations in the species' fitness may bring it into the
vicinity of one of its barriers, allowing it to crossover to a
different maximum.
This constitutes an evolutionary event in which the species changes
from one typical genotype to another.
For uncorrelated fluctuations, the expected time $\tau_{ij}$ to crossover
a barrier of height $f_{ij}$ will be given by an Arrhenius
equation of the form

\begin{equation}
\tau_{ij}\sim\exp(f_{ij}/f_{0})\:,
\label{e:se_tevolve}
\end{equation}

\noindent{}where the constant $f_{0}$ fixes the timescale and is analogous to
temperature.

Since the landscape is rugged, a species that escapes
from one maximum will soon become trapped by another,
and will find itself
surrounded by an entirely new set of barriers~$f_{ij}$.
In the limit \mbox{$f_{0}\rightarrow0$}, (\ref{e:se_tevolve})
implies that it will always be the species $i$ with the smallest
$f_{ij}$ in the system that evolves first,
and that the other species will have
moved no appreciable distance towards their own barriers
by the time this occurs.
Thus the ecosystem will consist of species
that infrequently hop between maxima at a rate that depends
upon the~$f_{ij}\,$, but are otherwise essentially static.
The evolution of species $i$ will alter the landscapes
of all those species $k$ linked to it in the ecosystem,
for instance via predator-prey
or host-parasite relationships.
Although each species $k$ will in general have to move
on their new landscapes before finding a new maximum,
it is a further approximation of the
theory that this can be ignored and only the
barriers $f_{kj}$ are affected.

The original Bak-Sneppen model is defined purely
in terms of the smallest barriers~$f_{i1}\,$,
which are arranged on a lattice in such a way that
interacting species occupy adjacent lattice sites.
The evolutionary process described in the previous
paragraphs then reduces to the dynamical interaction
between adjacent~$f_{i1}$.
The system is static until the site with the smallest
$f_{i1}$ evolves, when $f_{i1}$ and all of the $f_{k1}$
in adjacent sites $k$ are assigned new values.
The system is again static until another site evolves, and
so on.
It has been shown that the essential system behaviour is
insensitive to details such as the choice of probability
distribution for the $f_{i1}$ (Bak 1993, Paczuski 1996).
This {\em robustness}\/ relates to the universality
of the critical state, and is essential if
such a simple model is to faithfully describe real ecosystems.

The Bak-Sneppen model can be enhanced by a more detailed
consideration of the fitness landscapes and their interaction.
Three features absent from the original model will be considered here,
namely {\em speciation}, {\em extinction}\/ and {\em external
noise}. Each feature is described in general terms below
before the new model is fully specified.

\vspace{5mm}

\noindent{\em Speciation:}
Speciation occurs when two sub-populations reach
a state of reproductive isolation and should be
considered as separate species
(Maynard-Smith 1993, Ridley 1993).
For instance, two groups that are reunited after
prolonged geographical isolation may have evolved
so much in different directions that they are unable
to produce viable offspring.
Up until now, a species has been described as
simply occupying a region of genotype space.
More detailed analysis shows that a population
forms a ``cloud'' of points of roughly equal fitness
around the local maximum (Kauffman 1993).
Normally the whole population crosses over the same
barrier~$f_{i1}$, but if \mbox{$f_{i2}\approx f_{i1}$}
then it is possible that part of the population will
instead cross over the barrier~$f_{i2}$.
If this happens, the two subpopulations will move
to different maxima and a speciation
event will have occurred.
A simple criterion for speciation is to say that
species $i$ will branch into two subspecies if

\begin{equation}
f_{i2}-f_{i1}<\delta s
\label{e:se_spec}
\end{equation}

\noindent{}when it evolves to a new form,
where $\delta s$ is a constant parameter.
Further barriers could also be considered
to incorporate the simultaneous splitting into
three or more subspecies, but such events
will be very rare and are ignored here.

\vspace{5mm}

\noindent{\em Extinction:}
The system size would increase without limit if only speciation
were allowed, so some mechanism is required by which a species can
be made extinct and permanently removed from the system.
The original Bak-Sneppen model does not distinguish between
this form of extinction and {\em pseudo-extinction},
which is where a rapidly evolving species 
disappears from the fossil record if its intermediate forms
are not recorded.
What is required is some criterion for true extinction
defined purely in terms of the individual species' fitness landscapes,
analogous to~(\ref{e:se_spec}).
It is not clear how this may be achieved.
Instead, a heuristic approach is adopted here, which
is to say that
a species $k$ is made extinct if it is linked to
the species with the minimum barrier,
and has $f_{k1}$ greater than some threshold value.
This proves to be the simplest choice for which
the system size does not diverge.

\vspace{5mm}

\noindent{\em External noise:}
A fitness landscape is ultimately a function of the species itself,
the species with which it interacts,
and any factors external to the ecosystem,
so fluctuations in the inorganic environment can also cause the
fitness barriers to change.
Examples include local disturbances
such as volcanic eruptions or the formation of a new river,
to global events including meteor impacts and changes in the sea level.
The interactions between species have already been
incorporated into the model, but no allowance has yet been
made for these abiotic factors.
Continuing with the philosophy that changes in fitness
can be approximated by random variables,
external influences are assumed to alter
the fitness landscapes by an amount $O(\delta g)$ per unit time,
where $\delta g$ is a new parameter.
More precisely, every species in the system
will have their barriers altered by an amount

\begin{equation}
f_{ij}\rightarrow f_{ij}+\delta g_{ij}\:,
\label{e:se_noise}
\end{equation}

\noindent{}where the $\delta g_{ij}(t)$ are
uniformly distributed on $[-\frac{\delta g}{2},
\frac{\delta g}{2}]$ and uncorrelated in time.
External effects will occur on a separate timescale to the
evolutionary processes in~(\ref{e:se_tevolve}), but for
simplicity both timescales are fixed at the same constant
rate in this model.

\vspace{5mm}

It remains to be decided how interacting species are linked together.
The original Bak-Sneppen model placed the species on a regular
crystalline lattice in which interacting species occupy adjacent sites,
but this is not flexible enough to incorporate the addition of new
species to the system and is of no use here.
Real food webs have a much more involved structure,
and if the full range of interactions is allowed rather
than just links in the food chain, then it appears
that a great many species interact at least weakly
(Hall 1993, Caldarelli 1998).
Trying to model this would only serve to draw attention
away from the main motivation for the new model,
which is to allow a variable system size.
Instead, we adopt the mean field approach
in which each species $i$ interacts with $K-1$
other species $k$ chosen at random from the system.
The species $k$ are reselected at every time step.

The extended model can now be fully specified.
The ecosystem consists of $N(t)$ species labelled by
\mbox{$i=1\ldots N(t)$}.
Each species occupies the region around a local maximum
on a rugged fitness landscape, and is separated from
nearby maxima by barriers of various heights~$f_{ij}$.
The larger barriers can be ignored since they will
rarely contribute to the dynamics, but at least two
must be retained if speciation is to be incorporated.
Hence each species is defined by its two smallest
barriers $f_{i1}$ and~$f_{i2}$,
which are uniformly
distributed over the range [0,1]
and then ordered so that \mbox{$f_{i2}\geq f_{i1}$}.
The following steps (i)--(vi) are iterated for
every time step.

\vspace{0.2in}

\noindent{}(i) {\em Evolution:}
The smallest $f_{i1}$ in the system is
found and marked for evolution. It will move
to a new maximum in step (vi).

\vspace{0.05in}

\noindent{}(ii) {\em Speciation:}
If the single species marked for evolution has
$f_{i2}-f_{i1}<\delta s\,$,
then a new species is introduced to the system with
random barriers. $N\rightarrow N+1$.

\vspace{0.05in}

\noindent{}(iii) {\em Interaction:}
$K-1$ other species are chosen at random
from the remaining $N-1$ in the system.
They will be assigned new barriers in step (vi).

\vspace{0.05in}

\noindent{}(iv) {\em Extinction:}
If any the of the $K-1$ interacting species has
$f_{i1}>1$, it is removed from the system.
$N\rightarrow N-1$.

\vspace{0.05in}

\noindent{}(v) {\em External noise:}
Every barrier $f_{ij}$ in the system
is transformed according to (\ref{e:se_noise}),
and reordered if necessary.

\vspace{0.05in}

\noindent{}(vi) {\em New barriers:}
The species marked for evolution in step (i) and
the $K-1$ interacting species from step (iii) are
assigned new random barriers, ordered such that
\mbox{$f_{i2}\geq f_{i1}$}.

\vspace{0.2in}

With such specific definitions of general processes,
it is obviously important to check that the model is robust
to any arbitrary choices.
To test this, the simulations have been repeated with
various changes to the rules, and in no
case was any qualitative deviation observed.
For instance, different values for the extinction threshold
in step (ii) give the same behaviour, even if the threshold value
varies in time around a fixed mean.
Both $\delta s$ and $\delta g$ were chosen from
uniform, Gaussian and exponential distributions,
again with no apparent change in behaviour.
Since the model appears to be robust,
further discussion will be restricted to the
simple set of rules given in \mbox{(i)--(vi)}.
The threshold value for extinction was fixed at 1 to minimise
the number of new parameters.

Before continuing, it should be pointed out that the algorithm
presented in steps \mbox{(i)--(vi)} above is not exactly the same as
that described in our previous exposition of this work
(Head 1997).
This earlier model assumed that all $K$ species
``mutate'' (evolve) at every time step, whereas it is of course
just the species with the minimum barrier that evolves.
The corrected model studied here behaves in much the same
way as its previous incarnation,
except that the number of species is now only weakly dependent
on the connectivity $K$.
This is in agreement with data for real multispecies communities,
as discussed further in the next section.

\section{Results of numerical simulations}
\label{s:se_num}

The quantity of interest is the total system size $N(t)$.
This varies in a manner that depends upon the choice of
values for the parameters $K$, $\delta s$ and $\delta g$,
as described below.

\vspace{0.2in}

\noindent{}$\bullet$
\mbox{$\delta s=\delta g=0$}:
Steps (ii), (iv) and (v) never feature in the time evolution of
the system and the larger barriers $f_{i2}$ are redundant.
The $f_{i1}$ interact in the same way as the original
Bak-Sneppen model, the only difference being that
each $f_{i1}$ is the smaller of two uniform distributions
on [0,1] and so is distributed according to
\mbox{$P(f_{i1})=2(1-f_{i1})$}, \mbox{$f_{i1}\in[0,1]$}.
Since the model is robust to the choice of probability
distribution, this difference is not important.
$N(t)$ remains fixed at its initial value.

\vspace{0.05in}

\noindent{}$\bullet$
\mbox{$K=1$}:
There are no interactions, and the species that evolves
will always have \mbox{$f_{i1}<1$} so extinction
is impossible. \mbox{$N(t)\rightarrow\infty$}
if \mbox{$\delta s>0$} or remains fixed if \mbox{$\delta s=0$}.

\vspace{0.05in}

\noindent{}$\bullet$
{\em $K>1$, \mbox{$\delta s>0$} and \mbox{$\delta g=0$}}:
There is speciation but no extinction, $N(t)\rightarrow\infty$.

\vspace{0.05in}

\noindent{}$\bullet$
{\em $K>1$, \mbox{$\delta s=0$} and \mbox{$\delta g>0$}}:
There is extinction but no speciation, $N(t)\rightarrow0$.

\vspace{0.05in}

\begin{figure}
\centerline{\psfig{file=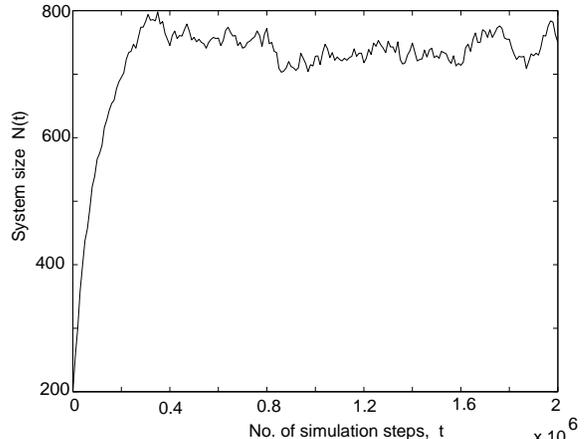,width=3in}}
\caption{Plot of $N(t)$ against $t$ starting from a system with
200 species. \mbox{$K=4$}, \mbox{$\delta s=0.008$}
and \mbox{$\delta g=0.02$}.}
\label{f:se_n_infty}
\end{figure}

\noindent{}$\bullet$
{\em $K>1$, $\delta s>0$ and $\delta g>0$}:
$N(t)$ fluctuates around some constant value $N_{\rm nat}$ which is
independent of the initial system size. An example
is given in Fig.~\ref{f:se_n_infty}.
Note that if \mbox{$\delta g\gg\delta s$},
$N_{\rm nat}$ is so small that statistical fluctuations will
eventually send \mbox{$N(t)\rightarrow0$} and the simulation is over.

\vspace{0.2in}

That $N_{\rm nat}$ should exist at all is by no means
obvious, since $N(t)$ does not explicitly appear in the
rules for speciation and extinction.
It exists because of the external noise of order $\delta g$,
which is just as likely to push
two barriers apart as to bring them
together and so does not affect the rate of speciation.
However, the noise acts asymmetrically
on barriers near the threshold for extinction,
tending to push species over this threshold
into the small tail corresponding to
those species that will be made extinct
when next selected.
Since every species is subjected to external noise
at every time step, the rate of extinction increases
with $N$ whilst the rate of speciation
remains roughly constant.
A steady state will be found when
these two rates balance.
This qualitative reasoning is confirmed
by the analysis in Sec.~\ref{s:se_anal}.

\section{Comparison to real multi\-spe\-cies com\-munities}
\label{s:se_real}

The parameters $\delta s$ and $\delta g$ are abstract quantities
defined purely in terms of the model, so it is not possible to
estimate their values for real ecosystems. Nonetheless it is
intuitively reasonable to assume that speciation and extinction
events are rare, and therefore both $\delta s$ and $\delta g$ should
be small.
The number of links per species $K-1$ has been measured for
real communities, and according to some studies is independent
of the system size
(Hall 1993, Kauffman 1993).
This is in agreement with numerical simulations of the model,
which shows only a weak dependence on $K$ from the range
\mbox{$K=2$} to \mbox{$K=16$}, as given in Table~\ref{t:se_num}
(at end of document).
There is a small peak around \mbox{$K\approx4$}, which also corresponds
to the most common value of $K$ observed in nature.
However, the data for real ecosystems is based on food webs
whereas the Bak-Sneppen approach considers all direct interactions
between species, so it is not clear how far this comparison can
be taken.

Turning to consider global ecosystems, the fossil record for
all marine organisms highlights the possibility of a statistical
steady state throughout much of the Palaeozoic era.
The total number of (families of) species fluctuates around
a roughly constant value up until the mass extinction at
the end of the Permian period, after which the system size increased
beyond its earlier levels and is still increasing today
(Benton 1995).
It could be conjectured that the new species that emerged after the
end-Permian extinction were on average either more likely to speciate,
or less susceptible to external noise, or both, which
should result in an increased system size according to the model.
The data for continental organisms is less clear and
if anything shows a continuing increase in diversity at varying rates.

\begin{figure}
\centerline{\psfig{file=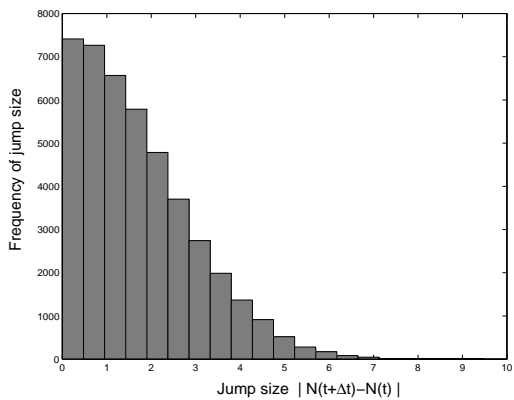,width=3in}}
\caption{Distribution of the absolute change in system size per
\mbox{$\Delta t=10^{3}$} time steps for \mbox{$K=2$}, \mbox{$\delta s=0.008$}
and \mbox{$\delta g=0.04$}. \mbox{$4\times10^{4}$} points were
sampled over 4 separate runs.}
\label{f:se_jump_size}
\end{figure}

The distribution of the magnitude of the change
in $N$ per unit time is Poisson to first order,
implying that the speciation and
extinction events are uncorrelated for \mbox{$N(t)\approx N_{\rm nat}$}.
However, the distribution is not {\em precisely}\/ Poisson,
which is presumably due to the tendency for $N$ to drift towards
$N_{\rm nat}$ when \mbox{$|\,N-N_{\rm nat}\,|$} is large.
An example is presented in Fig.~\ref{f:se_jump_size}.
That the distribution was not power law is disappointing,
but perhaps unsurprising given that the interactions between species
are randomised at every time step, making it difficult
for the system to self-organise.
It is possible that a spatially extended model might allow
for correlations to build up towards a critical state and a power
law to be recovered, but this must remain as speculation at present.

\section{Analytical derivation of $N_{\rm nat}$}
\label{s:se_anal}

It is possible to derive the dependence of $N_{\rm nat}$
on the parameters $K$, $\delta s$ and $\delta g$
by extending the mean field solution of the original
Bak-Sneppen model (Flyvberg 1993).
In theory this approach could give the exact solution,
since the interacting species are selected at
random in the new model and so it is mean field by definition.
However, the increased dynamical complexity means that
only the first order parameter dependence has be calculated.

\vspace{5mm}

\noindent{\bf Standard model with two barriers per species}

\noindent{}The original solution was based on a single barrier per species.
Before tackling speciation and extinction, it must first
be shown how the mean field approach can be modified
to handle pairs of barriers.
Define $p(x,y)\,{\rm dx\,dy}$ to be the
probability that a randomly selected species
has one barrier in the range [\,x,x+{\rm dx}\,)
and the other in [\,y,y+{\rm dy}\,).
Note that this refers to the barriers $f_{ij}$ {\em before}\/
ordering, so $x$ can be less that or greater than~$y$.
The probability for a species to have {\em both}\/
barriers greater than \mbox{$m=\min(x,y)$} is
represented by $Q(x,y)$, which is related to $p(x,y)$ by

\begin{equation}
Q(x,y)=\int_{m}^{\infty}\!\!\int_{m}^{\infty}p(x',y')\,{\rm dx'}{\rm dy'}\:.
\label{e:se_qdef}
\end{equation}

\noindent{}Since $p(x,y)=0$ for values of $x$ or $y$ outside
the range [0,1],
\mbox{$Q(x,y)=1$} for \mbox{$m\leq0$}
and \mbox{$Q(x,y)=0$} for \mbox{$m\geq1$}.
The species with the smallest barrier can be any of
the $N$ in the system, as long as all of the remaining
\mbox{$N-1$} species have larger barriers.
Hence $p_{\rm min}(x,y)$, the probability distribution
for the species with the smallest barrier, is given by

\begin{equation}
p_{\rm{min}}(x,y) = N\,p(x,y)\,Q^{N-1}(x,y) \:.
\label{e:se_pmin}
\end{equation}

At each time step, $p(x,y)$ will change by
an amount $\Delta p(x,y)$ which is given by the master equation

\begin{eqnarray}
&&\Delta p(x,y)  = -\frac{1}{N} p_{\rm{min}}(x,y)\nonumber\\
&&\mbox{}-\frac{K-1}{N-1}\left(
p(x,y)-\frac{1}{N} p_{\rm{min}}(x,y)\right)
+\frac{K}{N} \:.
\label{e:se_evolution}
\end{eqnarray}

\noindent{}The first term on the right hand side of (\ref{e:se_evolution})
accounts for the evolution of the species with the
lowest barrier, the second for the new barriers assigned
to the $K-1$ species with which it interacts,
and the third term handles the $K$ new pairs of
barriers.
In the statistical steady state,
$\Delta p=0$ and, using~(\ref{e:se_pmin}) and~(\ref{e:se_evolution}),

\begin{equation}
\frac{K}{N}-\frac{K-1}{N-1}p-\frac{N-K}{N-1}
pQ^{N-1}=0 \:.
\label{e:se_steady}
\end{equation}

\noindent{}The solution to (\ref{e:se_steady}) depends
upon the behaviour of $Q^{N-1}$ as \mbox{$N\rightarrow\infty$}
(Flyvberg 1993).
If $Q<1-O(1/N)$, then the term proportional
to $Q^{N-1}$ vanishes and

\begin{equation}
p(x,y)= \frac{K}{K-1} +O\left(\frac{1}{N}\right)\:.
\end{equation}

\noindent{}Conversely, if either $x$ or $y$ is so small
that $p(x,y)=O(1/N)$, then the second term in (\ref{e:se_steady})
will be $O(1/N^{2})$ and \mbox{$Q(x,y)=1+O(1/N)$}, so

\begin{equation}
pQ^{N-1}=\frac{K}{N}+O\left(\frac{1}{N^{2}}\right) \:,
\end{equation}

\noindent{}and hence from~(\ref{e:se_pmin}),

\begin{equation}
p_{\rm min}= K+O\left(\frac{1}{N}\right) \:.
\end{equation}

\noindent{}Each solution applies in different regions of the
$(x,y)$ plane, which, for large $N$, will be separated
by sharply defined boundaries.
These boundaries can be found by remembering that both $p$ and
$p_{\rm min}$ are probability distributions and normalise to one.
To first order in $1/N$, the full solutions are

\begin{eqnarray}
p(x,y)&\approx&\left\{
               \begin{array}{ll}
                     \frac{K}{K-1}  &
\mbox{$x$ and $y>1-\sqrt{\frac{K-1}{K}}$,} \\
                     0         &  \mbox{otherwise,}
               \end{array}
         \right.
\label{e:se_solp}\\
p_{\rm min}(x,y)&\approx&\left\{
              \begin{array}{ll}
                0  \hspace{0.2in} &   
\mbox{$x$ and $y>1-\sqrt{\frac{K-1}{K}}$,} \\
                K        &  \mbox{otherwise.}
              \end{array}
            \right.
\label{e:se_solpmin}
\end{eqnarray}

\noindent{}Hence the species with the smallest barrier will always
be found in the region where \mbox{$p_{\rm min}(x,y)\approx K$},
and its \mbox{$K-1$} interacting species will always come
from the region corresponding to \mbox{$p(x,y)\approx K/(K-1)$}.

\vspace{5mm}
\noindent{\bf Analysis for $\delta s$ and $\delta g$ non-zero but small}

\noindent{}When either \mbox{$\delta s>0$} or \mbox{$\delta g>0$},
the system size $N$ becomes a function of time
and the algebra quickly becomes prohibitive.
The simpler and more intuitive approach adopted here is to
initially ignore speciation and extinction altogether
and only incorporate the external noise of order~$\delta g$.
This leads to new solutions for $p(x,y)$ and $p_{\rm min}(x,y)$,
from which the rates of speciation and extinction
can be calculated even though they are no longer dynamically involved.
The natural system size $N_{\rm nat}$ is then the value of $N$
for which the two rates balance.
The analysis presented below assumes that $\delta g$ is small;
large values of $\delta g$ and $\delta s$ are considered
at the end of this section.

\begin{figure}
\centerline{\psfig{file=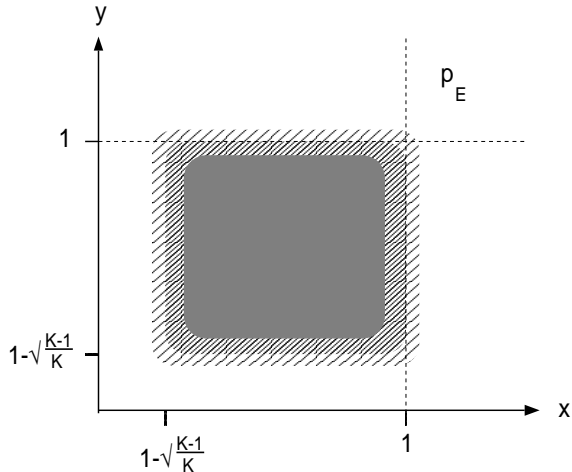,width=3in}}
\caption{Graphical representation of the effects of external noise
on the distribution of barriers~$p(x,y)$. Denser shading corresponds
to higher values of $p(x,y)$.
The region labelled ``$p_{\rm E}$'' refers to species with
both \mbox{$x>1$} and \mbox{$y>1$}, which are liable to extinction.}
\label{f:se_mfsol}
\end{figure}

The effect of the external noise
will be to perturb the solutions for $p$ and $p_{\rm min}$
given in (\ref{e:se_solp}) and~(\ref{e:se_solpmin}),
as schematised in Fig.~\ref{f:se_mfsol}.
The master equation~(\ref{e:se_evolution}) must be modified in two ways.
First, the external noise can cause barriers to move outside of
the range [0,1], so the range of possible $x$ and $y$ must be extended.
However, the barriers are still assigned values in the range
[0,1] and the term for new barriers must be altered accordingly.
Secondly, a term for the noise itself must be included.
The new steady state equation is

\begin{eqnarray}
\frac{K}{N}\,\theta(x)\,\theta(1\!-\!x)\,\theta(y)\,
\theta(1\!-\!y)-\frac{K-1}{N-1}p \nonumber\\
-\frac{N-K}{N-1}pQ^{N-1}
+\frac{\delta g^{2}}{24}\bigtriangledown^{2}p =0 \;\;,
\label{e:se_stdynoise}
\end{eqnarray}

\begin{equation}
\mbox{where}\;\;\; \theta(x) = \left\{ \begin{array}{ll}
      1  &  \mbox{if $x\geq0$,} \\
      0  &  \mbox{otherwise.}
  \end{array}
\right.
\end{equation}

\noindent{}The theta functions in the first term ensure
that new barriers lie in the range [0,1].
The last term on the right hand side of~(\ref{e:se_stdynoise})
accounts for the external noise, where $\bigtriangledown^{2}$ is the
Laplacian operator.
A full derivation of this term is given in the appendix.

\vspace{0.3in}

\noindent{\em Rate of extinction:}
Each of the $K-1$ random neighbours will be made extinct
if it has $x>1$ and $y>1$.
Thus the rate of extinction $k_{E}$ is given by

\begin{equation}
k_{E} =(K-1)
\int_{1}^{\infty}\!\!\int_{1}^{\infty}p(x,y)\,{\rm dx}\,{\rm dy}\:,
\label{e:se_ke}
\end{equation}

\noindent{}where the integral is over the region
$p_{\rm E}$ in Fig.~\ref{f:se_mfsol}.
Strictly speaking, the distribution in this
equation should be $p-\frac{1}{N}p_{\rm min}$, but
this distinction can be ignored for large $N$.
When both barriers are large, $Q^{N-1}\sim0$
and~(\ref{e:se_stdynoise}) can be simplified by the
transformation

\begin{eqnarray}
x\rightarrow x' & = & \alpha(1-x) \;\;, \\
y\rightarrow y' & = & \alpha(1-y) \;\;, \\
p\rightarrow p' & = & \frac{K-1}{K}p \;\;, \\
\alpha^{2} & = & \frac{48(K-1)}{\delta g^{2}N} \;\;,
\end{eqnarray}

\noindent{}to give

\begin{equation}
2\bigtriangledown'^{2}p'(x',y') =
p'(x',y')-\theta(x')\theta(y') \:.
\label{e:se_newvar}
\end{equation}

\noindent{}For either $x'$ or $y'$ negative, corresponding to
$x>1$ or $y>1$, the second term on the
right-hand side of~(\ref{e:se_newvar})
vanishes and the equation
can be solved by separation of variables.
Coupled with the boundary
conditions $p'(x',y')\rightarrow0$
for $x'\rightarrow-\infty$ or $y'\rightarrow-\infty$,
the solution is

\begin{equation}
p'(x',y') = c\,e^{\frac{1}{2}(x'+y')} \:,
\end{equation}

\noindent{}where $c$ is an arbitrary constant.
Whatever the value of $c$, it must be independent of
$K$ and $\delta g$ since these parameters do not appear
in (\ref{e:se_newvar}).
Transforming back into the original variables
gives the explicit parameter dependence,

\begin{equation}
p(x,y) = c\,\frac{K}{K-1}\,e^{-\frac{\alpha}{2}(x+y-2)}
\;\;\;\mbox{for $x>1$ and $y>1$} \:.
\end{equation}

\noindent{Substituting this into~(\ref{e:se_ke}) gives}

\begin{equation}
k_{E} \propto \delta g^{2}N\frac{K}{K-1} \;\;.
\end{equation}

\vspace{0.3in}

\noindent{\em Rate of speciation:}
It has not been possible to find a solution to
(\ref{e:se_newvar}) for \mbox{$x<1$} and \mbox{$y<1$}.
The variable separable solution does not behave correctly,
and other methods tried have been fruitless.
Instead, the $\delta g=0$ solution will be used as a first approximation.
The rate of speciation $k_{S}$ will be proportional to the density of
species with $|\,x-y\,|<\delta s$.
Since only the species with the minimum barrier can speciate,

\begin{equation}
k_{S}=\int\!\!\int \theta(\delta s\!-\!|x\!-\!y|)
\,p_{\rm min}(x,y) \,{\rm dx}\,{\rm dy} \:.
\label{e:se_k_s}
\end{equation}

\noindent{}Substituting the explicit expression
for $p_{\rm min}(x,y)$ from (\ref{e:se_solpmin}) into
(\ref{e:se_k_s}) gives

\begin{equation}
k_{S}\approx \delta s
\label{e:se_ks}
\end{equation}

\noindent{}for small~$\delta s$.
With $\delta g>0$, $p_{\rm min}(x,y)$ broadens
and so the real rate of speciation will decrease for larger~$\delta g$.

The value of $N_{\rm nat}$ can now be found up to
parameter dependence.
The rates of speciation and extinction balance when
\mbox{$k_{E}=k_{S}$}, and therefore

\begin{equation}N_{\rm nat}\propto\frac{\delta s}{\delta g^{2}}\:.
\label{e:se_ninfty}
\end{equation}

\noindent{}This implies that \mbox{$N_{\rm nat}\delta g^{2}/\delta s$}
should be roughly constant.
This quantity has been calculated from the numerical simulations
and is shown in Table~\ref{s:se_num}.
The agreement is good for variations in~$\delta g$,
but less so for $K$ and~$\delta s$. This most probably reflects the
first order approximation used in deriving $k_{S}$ in~(\ref{e:se_ks}).

\vspace{5mm}
\noindent{\bf Either $\delta s$ or $\delta g$ large}

\noindent{}For the sake of completeness, the equivalent expression to
(\ref{e:se_ninfty}) will now be derived for large $\delta s$
or~$\delta g$, although such values bear no relevance to
actual systems.
If $\delta s$ is large but $\delta g$ small, the system size rapidly
increases and with it the expected time a species will move
under the influence of external noise before being assigned new barriers.
Similarly, if both $\delta s$ and $\delta g$ are large,
then the system behaviour is also dominated by the external noise.
This is called the {\em noise dominated regime}.
If $\delta s$ is small but $\delta g$ large, $N_{\rm nat}$
becomes so low that fluctuations will eventually drive every species
in the system to extinction.

In the noise dominated regime, $p(x,y)$ will no longer
be just a small perturbation around the original solution
but will extend to large positive and negative values
in both the $x$ and $y$ directions.
Since the external noise is isotropic, $p(x,y)$ will
be symmetric about the $x$ and $y$ axes
and at most 1 in 4 species have both barriers in the
$p_{\em E}$ region.
Hence the rate of extinction will approach its upper
bound value of

\begin{equation}
k_{E} \approx \frac{K-1}{4} \:.
\label{e:se_kemax}
\end{equation}

\noindent{}When a barrier is assigned a new value in the range [0,1],
it undergoes an unbiased random walk until it is again assigned
a new value and brought back to near the origin.
The average number of steps in this walk will
be \mbox{$O(N/K)$} and, since the average step size is $O(\delta g)$,
an analogy with a one-dimensional random walker implies
that the total distance travelled will be
$O(\delta g\sqrt{N/K})$ (Papoulis 1991).
This gives the width of the barrier
distribution in both the $x$ and $y$ directions.
The number of species in the infinite
strip $|\,x-y\,|<\delta s$ is inversely proportional to the
width of $p(x,y)$, so the rate of speciation
is now given by

\begin{equation}
k_{S}\sim \frac{\delta s}{\delta
g}\sqrt{\frac{K}{N}}\:.
\label{e:se_ksnew}
\end{equation}

\noindent{}As $N$ increases, the rate of extinction
will remain roughly constant but now the rate of
speciation will decrease until a balance is
found at $N=N_{\rm nat}$.
From~(\ref{e:se_kemax}) and~(\ref{e:se_ksnew}),
the corresponding value of $N$ is

\begin{equation}N_{\rm nat} \propto \frac{1}{K}\left( \frac{\delta s}{\delta g}
\right) ^{2}\;\;.
\label{e:se_nnew}
\end{equation}

\begin{figure}
\centerline{\psfig{file=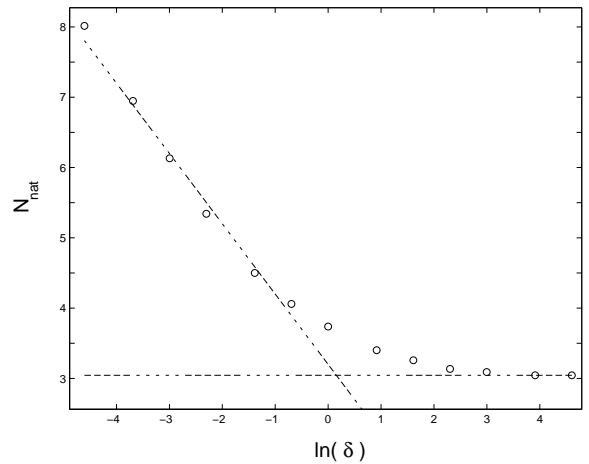,width=3in}}
\caption{Plot of $N_{\rm nat}$ versus
$\delta=\delta s=\delta g$ for \mbox{$K=2$},
demonstrating the crossover to the noise dominated regime.
Each point corresponds to a single run of $10^{6}$ time steps.
The dashed lines have slopes of $-1$ and~0.}
\label{f:se_crossover}
\end{figure}

\noindent{}A convenient way to display the crossover in behaviour
from small $\delta g$ to the noise dominated regime
is to consider $N_{\rm nat}$ as a function of
\mbox{$\delta=\delta s=\delta g$}.
According to (\ref{e:se_ninfty}) and (\ref{e:se_nnew}),
this should change from \mbox{$N\sim\delta^{-1}$} for small
$\delta$ to \mbox{$N\sim\delta^{0}$} for large~$\delta$.
Numerical results in support of this prediction are presented in
Fig.~\ref{f:se_crossover}.

\section{Discussion}
\label{s:se_disc}

In summary, we have postulated one possible way in which models of
macroevolution based on the Bak-Sneppen approach can be extended
to incorporate speciation, extinction and abiotic influences.
The speciation and extinction mechanisms are defined purely in
terms of each individual species' fitness landscape, irrespective of
the total number of species in the system.
Nonetheless, the total diversity fluctuates around a constant
value $N_{\rm nat}$, which was termed the {\em natural system size}\/
to stress that it was not arbitrarily chosen.

Although the proposed mechanism for speciation, {\em ie.}\/ a population
simultaneously crossing two different fitness barriers, seems
appealing, the extinction mechanism is far more heuristic
and somewhat unsatisfactory.
A better model might focus on trying to find a more plausible
means of extinction, defined in terms of the fitness landscapes.
For instance, the species chosen for evolution might be made
extinct if its fitness barrier is below some threshold value.
It may also be possible to place the model on a web structure,
and allow the connections themselves to be subject to
alteration whenever a species evolves to a new form.

The value of $N_{\rm nat}$ was found to be only weakly dependent
upon the average number of connections per species in the system,
in agreement with known data.
This leads us to hope that simple models such as ours may be able
to reproduce the essential behaviour of real ecosystems.
More realistic models consider the full fitness landscapes
rather than just the barriers, but the increased complexity
limits the system sizes that can be simulated (Kauffman 1993).
A practical step forward might be to reduce known biological principles
to simple rules that may be applied to global ecosystems.

\section*{Acknowledgment}

We would like to thank Prof.~Mark~Newman for useful discussions
concerning our model.


\section*{References}

\noindent{}{\bf Bak} P. and Sneppen K. 1993
{\em ``Punctuated equilibria and criticality
in a simple model of evolution''}  Phys. Rev. Lett. {\bf 71} 4083-4086

\noindent{}{\bf Bak} P. 1997
{\em ``How nature works:The science of
self-organized criticality''}\/ Oxford University Press

\noindent{}{\bf Benton} M. J. 1995
{\em ``Diversification and extinction in the history of life''}
Science {\bf 268} 52-58

\noindent{}{\bf Caldarelli} G., Higgs P. G. and McKane A. J. 1998
{\em ``Modelling coevolution in multispecies communities''}
preprint adap-org/9801003

\noindent{}{\bf Dawkins} R. 1982 {\em ``The extended phenotype''}
Oxford University Press

\noindent{}{\bf Flyvberg} H., Sneppen K. and Bak P. 1993
{\em ``Mean-field theory for a simple model of evolution''}
Phys. Rev. Lett. {\bf 71} 4087-4090

\noindent{}{\bf Hall} S. J. and Raffaelli D. G. 1993
{\em ``Food webs - Theory and reality''}
Adv. Ecol. Res. {\bf 24} 187-239

\noindent{}{\bf Head} D. A. and Rodgers G. J. 1997
{\em ``Speciation and extinction in a simple model of evolution''}
Phys. Rev. E {\bf 55} 3312-3319

\noindent{}{\bf Kauffman} S. A. 1993
{\em ``The Origins of Order''} Oxford University Press

\noindent{}{\bf Kramer} M. Vandewalle N. and Ausloos M. 1996
{\em ``Speciations and extinctions in a self-organising critical model
of tree-like evolution''}
J. Phys. I (France) {\bf 6} 599-606

\noindent{}{\bf Maynard-Smith} J. 1993 {\em ``The theory of
evolution''} Cambridge University Press

\noindent{}{\bf Newman} M. E. J. 1997
{\em ``A model of mass extinction''}
J. Theor. Biol. {\bf 189} 235-252

\noindent{}{\bf Paczuski} M. Maslov S. and Bak P. 1996
{\em ``Avalanche dynamics in evolution, growth and depinning models''}
Phys. Rev. E {\bf 53} 414-443

\noindent{}{\bf Papoulis} A. 1991
{\em ``Probability, random variables and stochastic processes''}
McGraw-Hill

\noindent{}{\bf Peliti} L. 1997
{\em ``An introduction to the statistical theory of Darwinian evolution''}
preprint cond-mat/9712027

\noindent{}{\bf Ridley} M. 1993 {\em ``Evolution''}
Blackwell Scientific Publications

\noindent{}{\bf Roberts} B. W. and Newman M. E. J. 1996
{\em ``A model for evolution and extinction''}
J. Theor. Biol. {\bf 180} 39-54

\noindent{}{\bf Sol\'{e}} R. V. and Manrubia S. C. 1996
{\em ``Extinction and self-organised criticality in a model of
large-scale evolution''} Phys.~Rev.~E~{\bf 54}
R42-45

\noindent{}{\bf Sol\'{e}} R. V., Manrubia S. C., Benton M. and Bak P. 1997a
{\em ``Self-similarity of extinction statistics in the fossil record''}
Nature {\bf 388} 764-767

\noindent{}{\bf Sol\'{e}} R. V. and Manrubia S. C. 1997b
{\em ``Criticality and unpredictability in macroevolution''}
Phys. Rev. E {\bf 55} 4500-4507

\noindent{}{\bf Wilke} C. and Martinetz T. 1997
{\em ``Simple model of evolution with variable system size''}
Phys. Rev. E {\bf 56} 7128-7131

\section*{Appendix}

In this appendix, the term for external noise that 
appears in~(\ref{e:se_stdynoise}) is derived.
Assuming that $\delta g$ is small, noise effects alone
will result in $p(x,y)$ taking the mean value of the
surrounding square with sides $\delta g$, that is

\begin{eqnarray}
&&\Delta_{\rm noise} \, p(x,y) = -p(x,y) \nonumber\\
&&\mbox{} + \frac{1}{\delta g^{2}}
\int_{x-\delta g/2}^{x+\delta g/2}
\int_{y-\delta g/2}^{y+\delta g/2}
p(u,v)\,{\rm du}\,{\rm dv}\:.
\label{a1}
\end{eqnarray}

\noindent{}Since $\delta g$ is small, $p(x,y)$ can be
expanded according to Taylor's theorem
as

\begin{eqnarray}
&&p(x+\delta x,y+\delta y)=
p(x,y) + \delta x\frac{\partial p}{\partial x}
\,+\,\delta y\frac{\partial p}{\partial y}\nonumber\\
&&\mbox{}\,+\,\frac{\delta x^{2}}{2!}\frac{\partial^{2}p}{\partial x^{2}}
\,+\,\delta x\delta y\frac{\partial^{2}p}{\partial x\partial y}
\,+\,\frac{\delta y^{2}}{2!}\frac{\partial^{2}p}
{\partial y^{2}}+\ldots
\end{eqnarray}

\noindent{}On substituting this into~(\ref{a1}), the terms
in $\delta x$, $\delta y$ and $\delta x\delta y$ integrate
to zero, leaving just the leading-order term

\begin{equation}
\Delta_{\rm noise} \, p(x,y) = 
\frac{\delta g^{2}}{24} \bigtriangledown^{2}p(x,y)
+O(\delta g^{3})  \:,
\label{a3}
\end{equation}

\noindent{}where $\bigtriangledown^{2}$ is the two-dimensional
Laplacian operator,

\begin{equation}
\bigtriangledown^{2} \equiv \frac{\partial^{2}}{\partial
x^{2}}
+\frac{\partial^{2}}{\partial y^{2}} \;\;.
\end{equation} 

\noindent{}The new term $\Delta_{\rm noise}\,p(x,y)$
is added to~(\ref{e:se_evolution}), the expression for $\Delta p(x,y)$
for when $\delta g=0$, to give the
total change in $p(x,y)$ at every timestep for $\delta g>0$.
Setting this to zero then gives the new steady
state equation~(\ref{e:se_stdynoise}).

\begin{table}
\begin{tabular}{|ccc|r|r|}
K & $\delta s$ & $\delta g$ & Numerical $N_{\rm nat}$
& $N_{\rm nat}\delta g^{2}/\delta s$ \\
\hline
\hline
2 & 0.008 & 0.02 & 625(19) & 31(1) \\
4 & 0.008 & 0.02 & 741(21) & 37(1) \\
6 & 0.008 & 0.02 & 717(22) & 36(1) \\
8 & 0.008 & 0.02 & 699(19) & 35(1) \\
16 & 0.008 & 0.02 & 615(17) & 31(1) \\
\hline
4 & 0.008 & 0.02 & 741(21) & 37(1) \\
4 & 0.008 & 0.04 & 187(12) & 37(2) \\
4 & 0.008 & 0.08 & 50(6)  & 40(5) \\
\hline
4 & 0.004 & 0.02 & 418(20) & 42(2) \\
4 & 0.008 & 0.02 & 741(21) & 37(1) \\
4 & 0.016 & 0.02 & 1257(25) & 31(1) \\
\end{tabular}
\caption{Observed values of the natural system
size $N_{\rm nat}$ for different $K$, $\delta s$ and~$\delta g$.
The standard deviation of the fluctuations are given in brackets,
which also serve as rough error bars.
The data has been averaged over at least three separate
rune of \mbox{$10^{6}-10^{7}$} timesteps each.
Note that the line for $K=4$, $\delta s=0.008$ and $\delta g=0.02$
appears three times to allow for easier comparison.}
\label{t:se_num}
\end{table}

\end{multicols}

\end{document}